MSc Research Project Report, Biology, Leiden University

# Adaptive Representations of Sound for Automatic Insect Recognition

Marius Faiß

Evolutionary Biology

Naturalis Biodiversity Center

Evolutionary Ecology

Darwinweg 2

2333 CR Leiden

The Netherlands

Supervisor: Prof. Dan Stowell

31/01/22 – 09/08/22

35 EC

Contact: Prof. Dan Stowell

Format: Plos Computational Biology

## Abstract:

Insects are an integral part of our ecosystem. These often small and evasive animals have a big impact on their surroundings, providing a large part of the present biodiversity and pollination duties, forming the foundation of the food chain and many biological and ecological processes. Due to factors of human influence, population numbers and biodiversity have been rapidly declining with time. Monitoring this decline has become increasingly important for conservation measures to be effectively implemented. But monitoring methods are often invasive, time and resource intense, and prone to various biases. Many insect species produce characteristic mating sounds that can easily be detected and recorded without large cost or effort. Using deep learning methods, insect sounds from field recordings could be automatically detected and classified to monitor biodiversity and species distribution ranges. In this project, I implement this using existing datasets of insect sounds (Orthoptera and Cicadidae) and machine learning methods and evaluate their potential for acoustic insect monitoring. I compare the performance of the conventional spectrogram-based deep learning method against the new adaptive and waveform-based approach LEAF. The waveform-based frontend achieved significantly better classification performance than the Mel-spectrogram frontend by adapting its feature extraction parameters during training. This result is encouraging for future implementations of deep learning technology for automatic insect sound recognition, especially if larger datasets become available.

## Background:

The insect order Orthoptera forms the animal clade with the most species capable of acoustic communication, with about 16,000 species using acoustic signals for sexual communication, and even more species displaying acoustic defensive signaling [1]. The sounds are produced by stridulation, where body parts are rubbed against each other to create audible vibrations, with one body part having a row of fine teeth and the other being equipped with a plectrum that sets the teeth into vibration [2–4]. Fossil records of homologous stridulatory sound production mechanisms from the Triassic period suggest that Orthoptera are the earliest known lineage that evolved complex acoustic communication [1]. Most of the 3200 species in the family Cicadidae produce sound by rapidly deforming tymbal membranes, producing series of loud clicking sounds that set the tymbals into resonance [5–7]. However, there are also species using stridulatory organs, like Orthoptera, or their wings to produce sound [6].

Hearing in Orthoptera is thought to be mostly focused on the time domain, with only crude analysis of fundamental frequency, mainly for distinction between conspecifics and predators [4,8]. The calls often contain amplitude modulation, resulting in chirps or more continuous trills and buzzes [4]. Duration and duty cycle of pulse and chirp sounds can be distinguished at high resolutions of less than 1 ms [9] as well as amplitude envelopes, meaning the length and decay time of individual pulses, and the length and pattern of pauses in between calls [4,8]. Sounds emitted by Orthoptera and Cicadidae are generally located in higher frequency ranges than most animal vocalizations, often achieved by frequency multiplication through the use of arrays of teeth or ribbed membranes that are set into vibration [4,7]. The frequency range and harmonic overtone structures of sounds these insects produce depend on the physical properties of the sound producing structures [2,4,10]. This is likely the cause for sonic differences between phylogenetic groups within Orthoptera, ranging from pure tones to more complex sounds with overtones, inharmonic sidebands and noise [2,4]. This allows the songs of Orthoptera species to show high stereotypy and to function as reliable taxonomic features [11]. The fundamental frequency of Orthoptera is inversely correlated with body size [2]. Temporal patterns of song play a role in the judgement of individual fitness or attractiveness, but can be disturbed by the presence of other singing conspecifics [8]. The fundamental frequencies of Orthoptera songs range from below 1 kHz up to 150 kHz [9,12]. Especially katydids (Orthoptera: Tettigoniidae) focus on ultrasonic sound production, with 66% of species producing ultrasonic frequencies higher than 20 kHz [2].

Declines in insect population numbers have been receiving wide attention in the scientific community as well as the public, but many of these reports only sample a small number of representative species or focus on limited geographic locations [13]. To implement effective conservation efforts, populations need to be monitored more closely and widely across species and geographic locations [13]. Acoustic monitoring methods focused on Orthoptera have been successfully used for detection of presence and absence of species, determining distribution ranges, evaluating quality and deterioration of habitats and detection of otherwise cryptic species [14], since they can function as indicator species [15]. Species that are difficult to observe due to physical size, camouflage, behavior, or habitat factors can be detected much more easily by the sounds they produce. Additionally, this method is mostly non-invasive, less elaborate than other common monitoring approaches and could be automated to a high degree [11]. Video monitoring in comparison, is highly dependent on lighting conditions and direct visual contact with the subjects, and consumes more energy as well as data storage [16].

However, there are also challenges and biases for acoustic monitoring. Some species only vocalize during short windows of time or season, which needs to be taken into account when planning surveys [14]. Detectability often varies with species-specific factors [17] or environmental factors and can make estimations of the number of individuals in a region difficult [18]. Certain features of vocalizations, especially loudness, but also frequency content or the position of the emitting individual in space can heavily influence the distance from which vocalizations can be detected [18]. Environmental features such as different types of vegetation and density also affect sound transmission differently and vary in absorption across the frequency range [18]. Other survey methods can have strong limiting factors just as much however, like lack of visual contact or recognition due to the specific environment or the species camouflage and more. But for determining distribution ranges or monitoring biodiversity and species richness, acoustic monitoring can be a useful alternative and might allow the detection of species or groups of animals that are hard to detect with other methods.

Insects are an especially difficult group to detect with conventional monitoring methods, mainly due to their small size, camouflage and cryptic lifestyles in often inaccessible and difficult environments such as tropical rainforests [11]. Previous attempts of identifying Orthoptera by their sounds have focused on using manual extraction of sound features such as carrier frequency or pulse rates [14]. These features have to be manually selected and their parameters defined before use for automatic classification. These features might not perform well in all situations however, for example when different species produce very similar sounds that do not significantly differ in the chosen feature dimensions or show strong variation of certain parameters. For example, ambient temperature during the recording can influence the frequency of Orthoptera song as a result of being poikilothermic organisms [19]. Orthoptera regulate their speed of muscular contraction with the ambient temperature during song production. This results in higher frequency sounds and especially increased pulse rates with higher temperatures in most Orthoptera [4,19].

Deep learning methods are a more recent promising approach for acoustic monitoring tasks, as they can classify acoustic signals with high accuracy and little to no manual pre-processing of the input data. Combined with sound event detection (SED), long-form field recordings can be classified without any manual extraction of features or relevant clips to be identified. There are however a number of challenges to overcome, some practical and some related to the specific species traits. For applying machine learning methods, large, diverse and balanced annotated datasets are needed to train and test the algorithms. If a dataset for a certain species is too small, the machine learning algorithms will learn only some of the kinds of variation included in that particular dataset and will therefore be less generalized [16]. For example, variations in the songs produced by a subset of a species population, or even single individuals, could be highly prioritized by an algorithm. This could lead to the misidentification of conspecifics that do not have these specific attributes. Other kinds of variation in the data could come from external factors, such as the environmental sound scene or sound absorption, background noise or distance of the sound source from the recorder [16].

There are however methods to expand and further diversify already existing datasets, by simulating the factors that can lead to realistic variation in the data. This method, called data augmentation, deforms aspects of annotated data samples to increase the diversity and quantity of training data sets without having to collect and annotate new data [20]. An important requirement is that the meaning of the annotated data is not changed and still relates to the assigned label [20,21]. Ideally, this should allow the neural network to generalize more on unseen data with variation that was not present in the non-augmented datasets. A common approach is to shift recordings in time or frequency, or erase parts of the frequency spectrum or timeline [22]. Artificial noise or realistic background noise that could be present at recording sites can be added to recordings to train the network for more challenging identification tasks that it might encounter with real world data. The application of impulse responses from real environments can simulate distance from the recorder and reflection or sound absorption characteristics of the environment [23,24]. This approach to data augmentation is not yet commonly used but could be a realistic and useful method to add to the standard audio augmentation toolkit. It could be especially useful if the datasets used for training are sourced from professional recordists or from stationary recording devices that can achieve high quality, focused recordings from close distances, but the recordings that are supposed to be classified will mostly come from amateur recordists that can not get into close proximity of the sound sources or do not have access to highly directional recording devices. Data augmentation methods have been shown to potentially affect different label classes of data differently [20]. However, simulating deformations or variation that could realistically occur with real-world data that is supposed to be classified by a network should generally be beneficial for classification tasks and are commonly used.

For building a tool that can be applied to field recordings with the goal of detecting and identifying all sound producing species that are present, a sound event detection and classification network would be ideal. This way, recordings would not have to be pre-processed by extracting clips with vocalizations, removing background noise, etc. A network like this could also ideally handle overlapping vocalizations by multiple individuals or species, and noisy situations. This is a complex task however, requiring intense training and tweaking of the network parameters.

Before an audio recording can be fed into a neural network to be analyzed, the high-resolution waveform has to be reduced to a feature space that can be processed and interpreted by a neural network [25,26]. The common approach for audio classification tasks has historically been inspired by the human perception of frequency and loudness. This is in part due to the focus of many of the early audio classification tasks that were heavily researched, like speech or language recognition, or music-based analysis tasks [26]. All the relevant acoustic information for these tasks is contained in and optimized for human auditory perception, or vice versa. Humans experience frequency and loudness on non-linear scales [27]. Linear changes in frequency towards the lower frequency spectrum generally sound more severe, while the same difference in frequency applied to a higher register can be undetectable to the human ear. In compressing the spectral energy of a signal for analysis in a neural network, these characteristics of human perception are applied with the use of Mel-filter banks.

First, the input audio waveform is transformed into a spectrogram using the short-time Fourier transform (STFT), dissecting the signal into pure sine-wave frequencies and their respective energies [25,27]. Then, the Mel-filter banks are applied, consisting of triangular bandpass filters, spaced along a logarithmic scale over the sampled frequency spectrum. These filters pool the energy of all frequencies that lie within their range, using a windowing function. This reduces the resolution from a high sample rate down to a number of frequency bins that can be easily analyzed. Following this, loudness compression is applied, also based on the non-linearity of human hearing [25], resulting in a Mel-spectrogram, that can essentially be treated like an image by a neural network. These processing methods, especially the filter banks, rely on hand-crafted parameters, that may not relate in any way to the sounds to be analyzed in a specific task, or to the inner workings of the network that is used. Logarithmic scaling for example, results in high spectral resolution in lower frequency ranges, but groups together larger and larger frequency ranges in higher registers, thereby potentially obscuring relevant high-frequency information, but also unnecessarily focusing on lower frequency bands when they do not necessarily contain relevant information (Fig. 1).

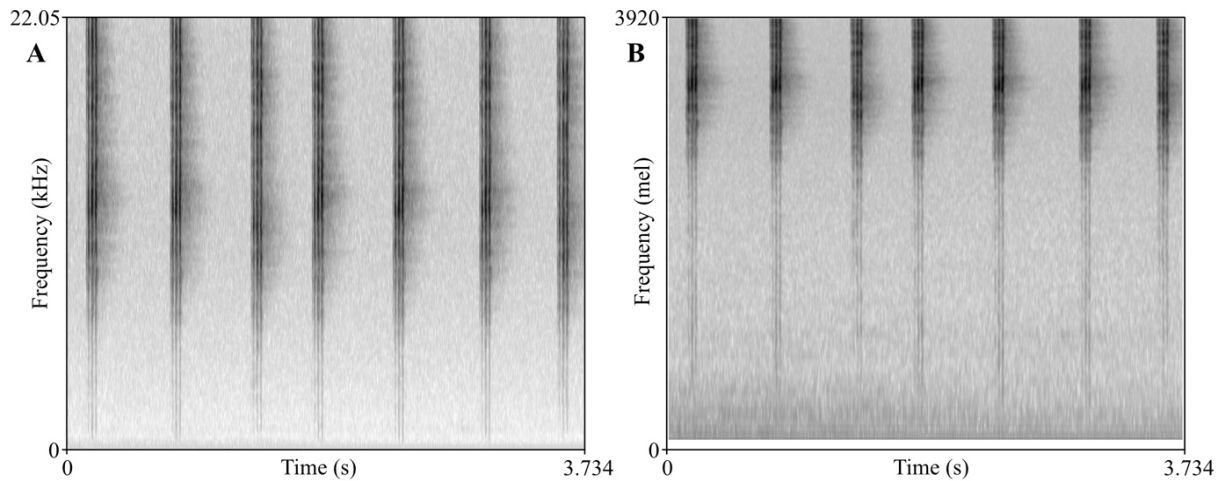

Fig. 1: Two spectrograms of the same recording of *Pholidoptera griseoaptera*. Spectrogram A displays the frequency axis linearly in Hz. Spectrogram B uses the Mel frequency scale, which compresses the frequency axis to show higher resolution in lower frequency bands than in higher bands, mimicking the human perception of frequency. Both spectrograms display the same spectrum of frequencies. Due to the mostly high-frequency information and empty low frequencies in this recording, the Mel spectrogram B obscures a large amount of information compared to the linear spectrogram A.

Insect sounds are not generated using a source-filter mechanism as in mammals or birds, but with stridulatory or tymbal mechanisms that create a different structure of frequencies and overtones [2–7]. Generally, insect sounds are much higher in frequency than most mammal or bird sounds, with many species producing ultrasonic sounds [5,9,12]. This emphasis on high-frequency sounds, sometimes entirely and far outside of the human hearing range (~20 Hz - 20 kHz) could have an impact on the performance of audio classification networks, depending on their approach. It is likely that the Mel-filter bank approach based on human perception might not be optimal to recognize and discriminate between subtle differences in high frequencies for many insect sounds.

Adaptive, waveform-based methods such as LEAF [25] could potentially optimize their extraction of audio features to better fit insect sounds. The LEAF frontend allows the adjustment of filter frequency and bandwidth, normalization parameters, and time-pooling parameters during training to adapt to the data [25]. This frontend has been evaluated on a diverse set of audio classification tasks involving human-centric sound such as language, music, emotion, speaker recognition and more, and has shown improved performance over the standard Mel spectrogram approach in many cases [25]. But so far, it has not been evaluated on classification tasks involving sound sources that are less fit to the human perception of sound. For uses like insect species recognition that are much higher pitched and structured differently than human sounds, this frontend could have and even higher advantage. It could gain an advantage over the fixed Mel frontend by learning increasing spectral resolution in higher frequency ranges, selecting and focusing on meaningful frequency bands that are otherwise pooled together, and learning how to ideally pool and compress these bands individually. Accordingly, the high resolution in lower frequency ranges that is present in Mel-filter bank approaches could be reduced or completely omitted, since it is rarely present in insect sounds (< 1 kHz) [9].

The performance of adaptive representations has recently been evaluated on various human-centric audio classification tasks, as well as bird classification, and has been shown to increase accuracy in most tasks and overall [25]. The potential of deep learning methods for insect sound classification has not been studied extensively yet, especially their performance with adaptive frontends and extended sample rates/frequency ranges. In this project, the performance of two different machine learning approaches will be tested in single task classification of monophonic insect sound recordings, with only one species or individual present at once. Complicating environmental conditions like distance from the recorder or background noise will be introduced by data augmentation methods to increase the diversity of the data set and improve the generalizability of the networks. The goal is to explore the potential for using deep learning methods to classify Orthoptera and Cicadidae sounds and to evaluate the potential advantage of adaptive frontends for feature extraction of non-human, high-frequency sounds.

# Methods:

I compared the performance of two different deep learning frontends with the same neural network backend structure to make the classification performances consistent and comparable. I compared the classic Mel-spectrogram frontend to the adaptive and waveform-based frontend LEAF, which is initialized on similar parameters as the Mel frontend, but can be adjusted during training [25]. As a backend classifier, a convolutional neural network optimized for audio classification was implemented and adapted [28].

The datasets were sourced from collections of insect sound researchers (Baudewijn Odé, unpublished, 177 recordings of Orthoptera; Ed Baker, unpublished, 547 recordings of Cicadidae). All files were manually inspected and files with strong noise interference or with sounds of multiple species were removed. A high number of files in one of the datasets included live audio comments made by the recordists during the beginning of the recording. The last ten seconds of audio were automatically extracted to be included in the dataset since these were unaffected by the comments in most cases. Only species with at least four usable recordings were included in the final dataset. Overall, 32 species were selected, with 335 files and a total recording length of 57 minutes and four seconds (Table 1). Between species, the number of files ranges from four to 22 files and the length from 40 seconds to almost nine minutes of audio material for a single species. The files range in length from less than one second to several minutes. All original files were available with sample rates of at least 44.1 kHz or higher but were resampled to 44.1 kHz for consistency. The sounds recorded in this dataset are often wide-band noisy emissions covering broad ranges of the frequency spectrum. Most species emit their loudest frequencies upwards of 4-5 kHz, but often reach higher than the maximum frequency of 22.05 kHz included in the data due to the sample rate. Very few recordings contain significant insect sound emissions below 1 kHz. The lowest fundamental frequencies in this dataset reach down to approximately 300 Hz, emitted by the species *Kikihia muta*. Frequencies in this range are extremely low for insects to produce and are likely an artifact of the proximity effect picked up by the recorder at a very close distance, accentuating low frequencies. Many species of Cicada contract their tymbal muscles at lower frequencies, but curved ribs on the tymbal membranes produce series of clicks at much higher rates, acting as frequency multipliers that produce the loudest frequencies [7]. It is not likely that frequencies this low radiate far from an insect and form a significant part of their sound signature.

Table 1: Number of files (n) and total length of recordings (min) per species included in the dataset. 335 files from 32 species with a total recording length of 57 minutes and four seconds were selected from two different source datasets (Orthoptera dataset by Baudewijn Odé and Cicadidae dataset by Ed Baker).

| **Baudewijn Odé - Orthoptera** | | | **Ed Baker - Cicadidae** | | | | | |
|---|---|---|---|---|---|---|---|---|
| **Species** | **n** | **min** | **Species** | **n** | **min** | **Species** | **n** | **min** |
| Chorthippus biguttulus | 20 | 3:43 | Azanicada zuluensis | 4 | 0:40 | Platypleura divisa | 6 | 1:00 |
| Chorthippus brunneus | 13 | 2:15 | Brevisiana brevis | 5 | 0:50 | Platypleura haglundi | 5 | 0:50 |
| Gryllus campestris | 22 | 3:38 | Kikihia muta | 6 | 1:00 | Platypleura hirtipennis | 6 | 0:54 |
| Nemobius sylvestris | 18 | 8:54 | Myopsalta leona | 7 | 1:10 | Platypleura intercapedinis | 5 | 0:50 |
| Oecanthus pellucens | 14 | 4:27 | Myopsalta longicauda | 4 | 0:40 | Platypleura plumosa | 19 | 3:09 |
| Pholidoptera griseoaptera | 15 | 1:54 | Myopsalta mackinlayi | 7 | 1:08 | Platypleura sp04 | 8 | 1:20 |
| Pseudochorthippus parallelus | 17 | 2:01 | Myopsalta melanobasis | 5 | 0:43 | Platypleura sp10 | 16 | 2:24 |
| Roeseliana roeselii | 12 | 1:03 | Myopsalta xerograsidia | 6 | 1:00 | Platypleura sp11 cfhirtipennis | 4 | 0:40 |
| Tettigonia viridissima | 16 | 1:34 | Platypleura capensis | 6 | 1:00 | Platypleura sp12 cfhirtipennis | 10 | 1:40 |
| | | | Platypleura cfcatenata | 22 | 3:34 | Platypleura sp13 | 12 | 2:00 |
| | | | Platypleura chalybaea | 7 | 1:10 | Pycna semiclara | 9 | 1:30 |
| | | | Platypleura deusta | 9 | 1:23 | | | |

For training and evaluating the two frontends, the dataset was split into three parts, the training, validation and test sets [16]. The training set is fed into the network and used to learn how to adjust the network parameters to better fit the data. The validation set is not used for learning directly, but to get a performance estimate on data that has not been used to train the network. It allows tweaking of the hyperparameters that control the learning process. The test set is used only for final evaluation and contains data that is completely unseen by the model and has not been used for parameter tweaking, simulating real-world data. A common split ratio for the three datasets is 70% training set, 10% validation set and 20% test set. Due to the low number of files in some classes in the original dataset, the split into the three datasets was done for all classes individually to ensure that each class is represented in all datasets. The resulting split amounts to 210 files being used for training (62.7% of the data), 51 files for validation (15.2%) and 74 files for testing (22.1%) (Fig. 2).

Since the files were different in length, they had to be divided into segments of a fixed length that can be fed into the network. A length of five seconds was chosen, as most calls were either short and rhythmical or long and static. Repeating sequences of longer than five seconds were not observed in the dataset, therefore it was assumed that a length of five seconds would not eliminate species-specific rhythmic characteristics in the calls.

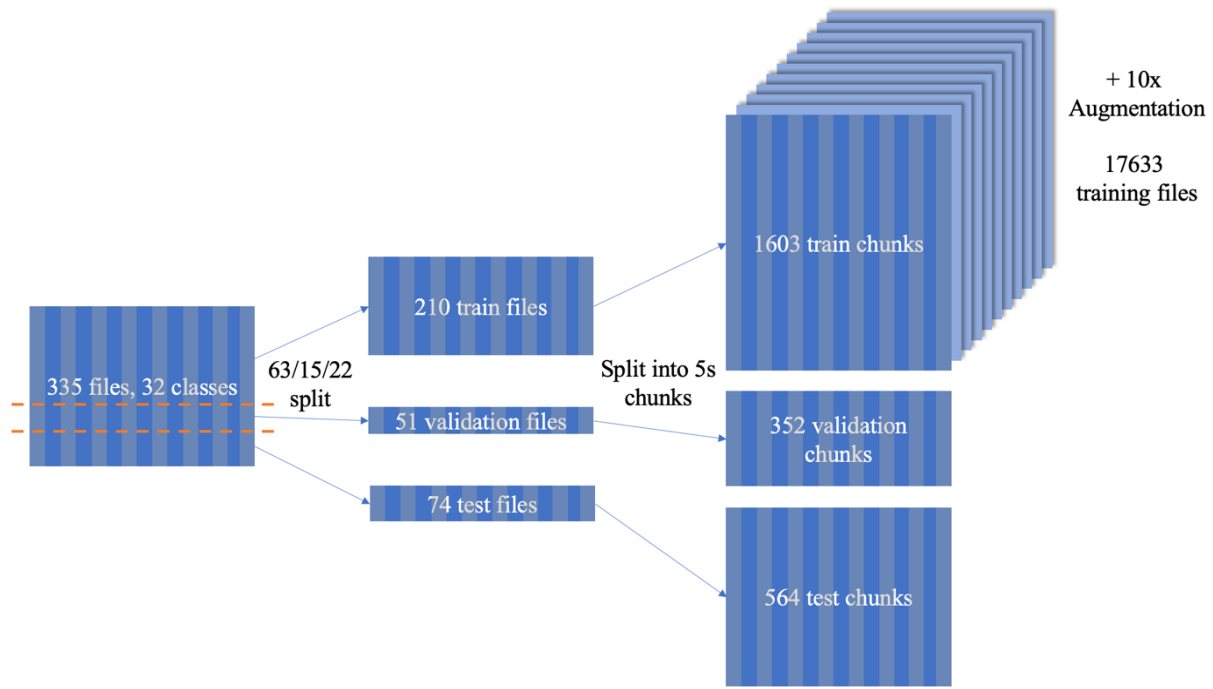

Fig. 2: Division of the original dataset, containing 32 species classes, into train, validation and test sets. All files were split into overlapping five second chunks of audio. The training set size was artificially increased with the addition of 10 audio augmentation treatments.

This truncation step was used as an opportunity for data augmentation. Files shorter than five seconds were looped until they reached the desired length. Files longer than five seconds were sequentially spliced into chunks of five seconds, with an overlap of 3.75 seconds. When the splitting window reached the end of a file, the beginning of the recording was wrapped around to extend the chunk to five seconds, as long as the minimum remaining time of a chunk was at least 1.25 seconds. This resulted in a more than seven-fold increase in file numbers, and in a more diverse dataset than if the files had only been truncated to a fixed length (Fig. 2). Some recordings of the species *Pholidoptera griseoaptera* contained long periods of silence in between calls and were manually split into chunks to avoid automatically generating empty chunks.

The training dataset was additionally expanded with ten generations of audio augmentations using the python package "audiomentations" (Fig. 3). The processing steps included "FrequencyMask", which erases a band of frequencies around a random center frequency, with bandwidth as a parameter that can be randomized within a defined range (Fig. 3). The bandwidth range was set between 0.06 and 0.22. This augmentation step was applied with a chance of 50%. After frequency masking, the signal was mixed with gaussian noise, using the "AddGaussianSNR" function. The ratio of signal to noise was randomized between 25 and 80. This ratio was tuned to range from barely noticeable addition of noise to heavy noise disturbance that did however not completely obscure all audio information in already noisy recordings. This was applied to every file. After mixing with noise, the files were augmented with impulse responses (IRs) recorded in natural outside settings (Fig. 3). The IRs were selected from a dataset of recordings made in various locations at high sample rates [29]. Eleven IRs from three different outside locations (two forest locations, one campus location) were selected from this dataset and randomly applied during augmentation with a chance of 70% (Fig. 3). The IR-processed files were mixed with their original version at random mix ratios to achieve additional variation in the severity of the effect (Fig. 3).

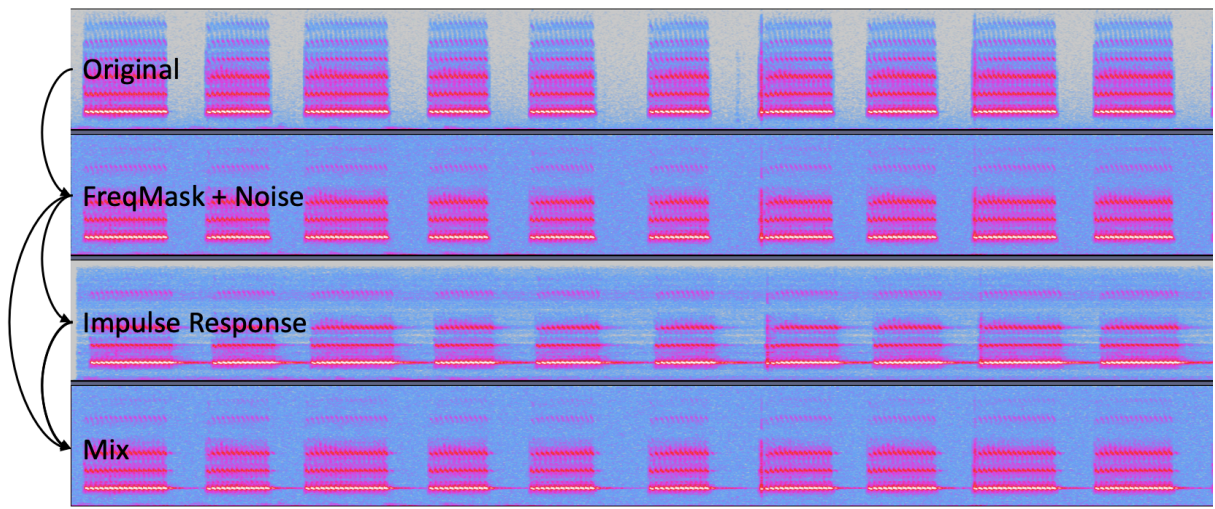

Fig. 3: Example of the data augmentation workflow used on the training set. Randomized frequency masking and noise mixing were applied to the files. Then an impulse response was selected randomly and applied at a randomized mix ratio.

The frontends that were compared are the conventional Mel spectrogram included in the python package torchaudio (MelSpectrogram) and the adaptive, waveform-based frontend LEAF [25]. The Mel spectrograms were generated based on the audio waveforms before the files were input into the network. When using the LEAF frontend, the full waveforms were directly input to the network and then processed by the frontend, since many of its parameters like filter frequency and bandwidth, per-channel compression and normalization, and lowpass pooling can be learned and therefore need to be part of the network to benefit from gradient descent learning. The initialization parameters of the two frontends were defined as similarly as possible to create a fair comparison. The files were imported at a sample rate of 44.1 kHz.

They were transformed from an input shape of [1; 220500] (one channel mono audio; 44.1 kHz for five seconds) to a representation shape of [1; 64; 1500] by the frontends, with 64 filter bands on the frequency axis and 1500 steps dividing the time axis. The window length was set at twice the length of the stride for both frontends (Mel stride: 147, Mel window size: 294; LEAF stride: 3.335; LEAF window size: 6.67). The filter bank used in the LEAF frontend was initialized on the same scale as the Mel frontend, between 0 and 22.05 kHz. The inputs were combined into batches of 14 and fed into the network.

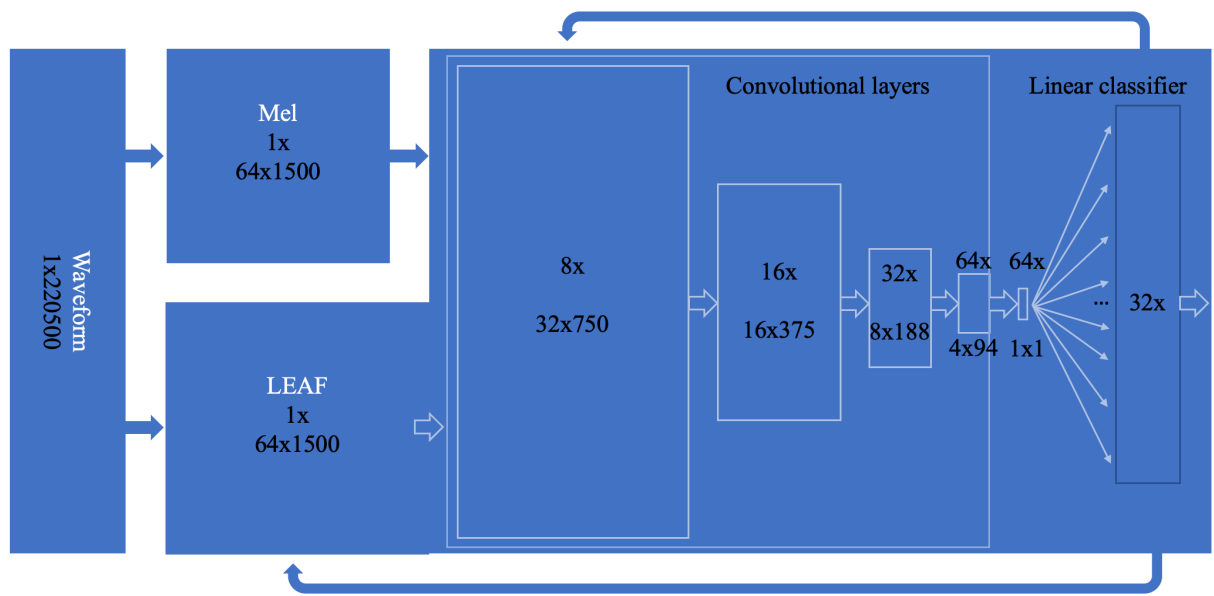

Fig. 4: The frontend and network architecture used in this experiment. The Mel frontend was implemented as a processing step before the network, while the LEAF frontend is integrated into the network where its parameters are optimized by the learning process during training.

The network backend was adapted from a convolutional neural network created using pyTorch that was optimized for audio classification [28]. It consists of four convolutional layers (Conv2d) with rectified linear units (ReLU) and batch normalization (BatchNorm2d). After the convolutional layers, the feature maps were pooled (AdaptiveAvgPool2d) and flattened, and finally input into a linear layer (Linear) that returns a prediction value for each of the 32 classes. The highest prediction value was picked as the final predicted class for each training example. To avoid overfitting of the network on the training dataset, dropout was implemented on the final linear layer (dropout rate of 0.4), as well as L2 regularization of the weights (weight decay of 0.001). Overall, the model contains 27,344 trainable parameters that are adjusted during the training phase, with the inclusion of the LEAF frontend (Suppl. Table 3).

The training process employed early stopping, which evaluates the network performance after each epoch by running an inference step on the validation set. The loss value of the validation set is used to estimate how well the network will perform on the test set during final evaluation. Each time the validation loss decreases, the current network state is saved. If the validation loss does not decrease any further for eight epochs the training is stopped and the final test evaluation is performed on the last saved network state from eight epochs earlier. The accuracy of the two approaches was determined by the percentage of correctly classified items in the test set, as well as the f1-score, precision and recall [16]. Due to the randomness included in the training process due to dataset shuffling and network initialization, the training and evaluation outcomes can vary substantially between runs using the exact same parameters and datasets. To achieve a stable and comparable result, both models were computed five times and the best performing models were selected based on the lowest validation loss value (Table 2).

## Results:

Table 2: Test and validation scores for all five model training iterations using both frontends. The best performing iterations were selected based on the lowest validation loss scores (highlighted in green).

| | Test | | | | Validation | | |
|---|---|---|---|---|---|---|---|
| **Model iteration** | **Accuracy** | **F1-score** | **Recall** | **Precision** | **Accuracy** | **Loss** | **Epoch** |
| Mel 1 | **0.67** | 0.56 | 0.58 | 0.64 | 0.65 | **1.37** | 26 |
| Mel 2 | 0.62 | 0.52 | 0.53 | 0.61 | 0.61 | 1.49 | 9 |
| Mel 3 | 0.66 | 0.55 | 0.58 | 0.62 | 0.60 | 1.44 | 14 |
| Mel 4 | 0.57 | 0.47 | 0.49 | 0.61 | 0.57 | 1.68 | 7 |
| Mel 5 | 0.60 | 0.48 | 0.52 | 0.52 | 0.57 | 1.56 | 18 |
| Mel Avg. | 0.62 | 0.52 | 0.54 | 0.60 | 0.60 | 1.51 | 14.8 |
| LEAF 1 | 0.76 | 0.66 | 0.67 | 0.70 | 0.66 | 1.24 | 23 |
| LEAF 2 | 0.76 | 0.69 | 0.71 | 0.73 | 0.71 | 1.16 | 33 |
| LEAF 3 | **0.78** | 0.69 | 0.70 | 0.73 | 0.76 | **1.00** | 30 |
| LEAF 4 | 0.77 | 0.66 | 0.68 | 0.67 | 0.71 | 1.25 | 26 |
| LEAF 5 | 0.59 | 0.61 | 0.60 | 0.69 | 0.61 | 1.40 | 17 |
| LEAF Avg. | 0.73 | 0.66 | 0.67 | 0.70 | 0.69 | 1.21 | 25.8 |

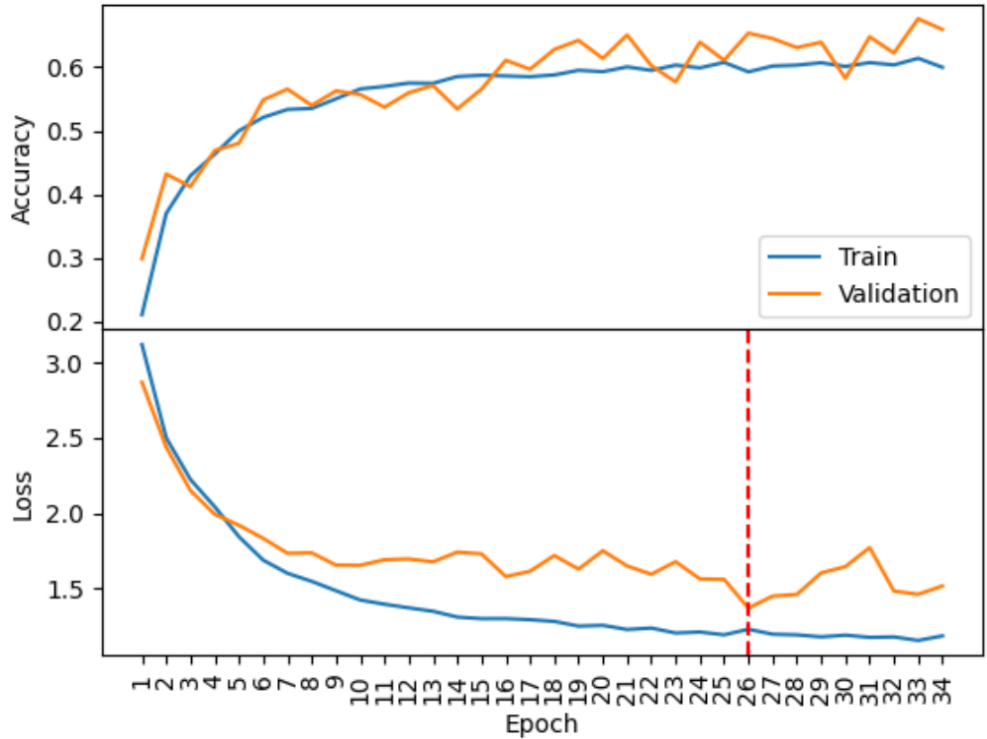

Fig. 5: Training progress of the best iteration using the Mel frontend, showing training and validation accuracy and loss. The training was stopped after 26 epochs, when the validation loss reached its lowest value of 1.37 and the validation accuracy 65%.

The results from training and evaluating five iterations of the Mel spectrogram frontend ranged between 57% and 67% classification accuracy on the test set and validation loss values of 1.37 and 1.68 (Table 2). Performance of the LEAF frontend ranged from 59% test set classification accuracy to 78%. The best validation loss score of a LEAF iteration was reached with a value of 1.00, while the lowest score was only slightly worse than the best score reached with the Mel frontend at 1.40 (Table 2). Four out of five test accuracy scores of the LEAF frontend fall between 76% and 78%, with the fifth iteration being a clear outlier in test set classification performance. The variation in test set classification between the Mel iterations is more varied. When looking at the additional performance metrics F1-score, recall and precision, even the worst performing LEAF iteration outperforms all of the Mel iterations (Table 2). On average, the LEAF iterations trained for eleven epochs longer than the Mel iterations (Table 2).

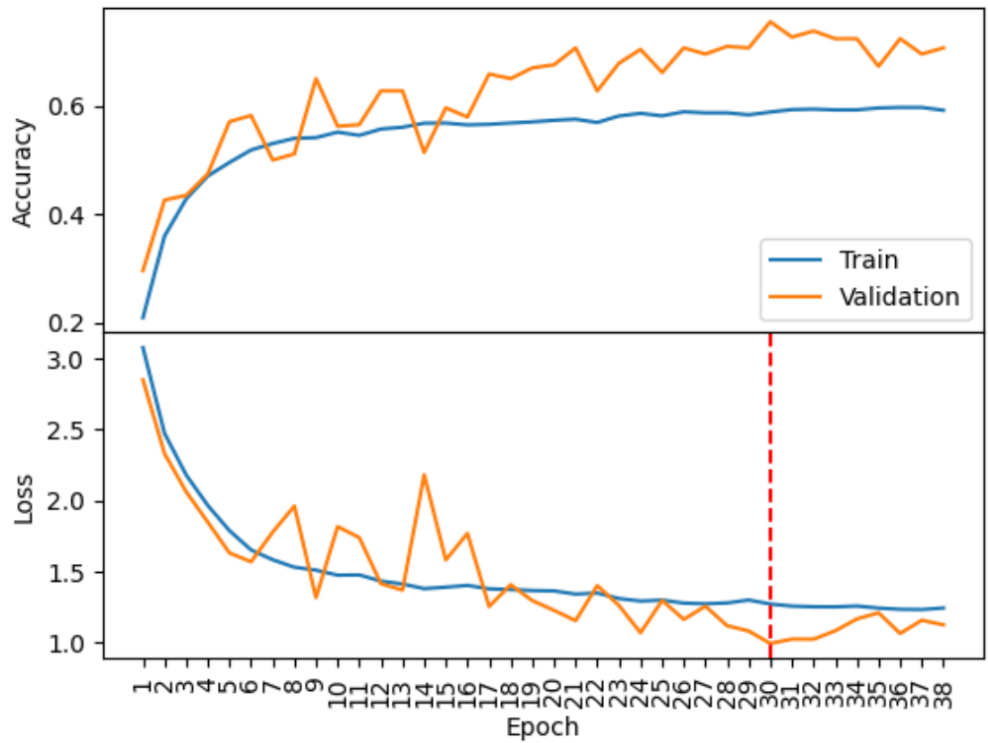

Fig. 6: Training progress of the best performing LEAF frontend iteration, showing training and validation accuracy and loss. The training was stopped after 30 epochs, when the validation loss reached its lowest value of 1.00 and the validation accuracy reached 76%.

The best performing Mel frontend iteration based on the lowest validation loss value achieved a minimum loss of 1.37 and a validation accuracy of 65% after 26 epochs of training (Fig. 5, Table 2). The final evaluation on the test set using the model at its optimal state resulted in a classification accuracy of 67%. Additional performance measures include the F1 score at 0.56, a recall of 0.58 and precision of 0.64 (Table 2). The best performing LEAF iteration reached its minimum validation loss value of 1.00 and validation accuracy of 76% after 30 epochs of training (Fig. 6, Table 2). The test set evaluation using the model at its optimal state resulted in a classification accuracy of 78%. The F1 score reached 0.69, the recall 0.70 and the precision 0.73 (Table 2).

The confusion matrix visualizing all classifications on the test set using the Mel frontend shows a visible diagonal line of correct classifications, but also many incorrect classifications deviating from the line (Fig. 7). The majority of misclassifications lie within the two biggest genera represented in the dataset, *Myopsalta* and *Platypleura* (5 and 14 species respectively, of 32 in total; Table 1). Species from these genera were most often misclassified as other members from the same genus. Especially the species *M. leona* caused many misclassifications within its genus, despite being entirely correctly classified itself. Within *Platypleura*, the species *P. plumosa* and *P. sp12 cfhirtipennis* were especially often wrongfully assigned as labels to samples of other members of the same genus.

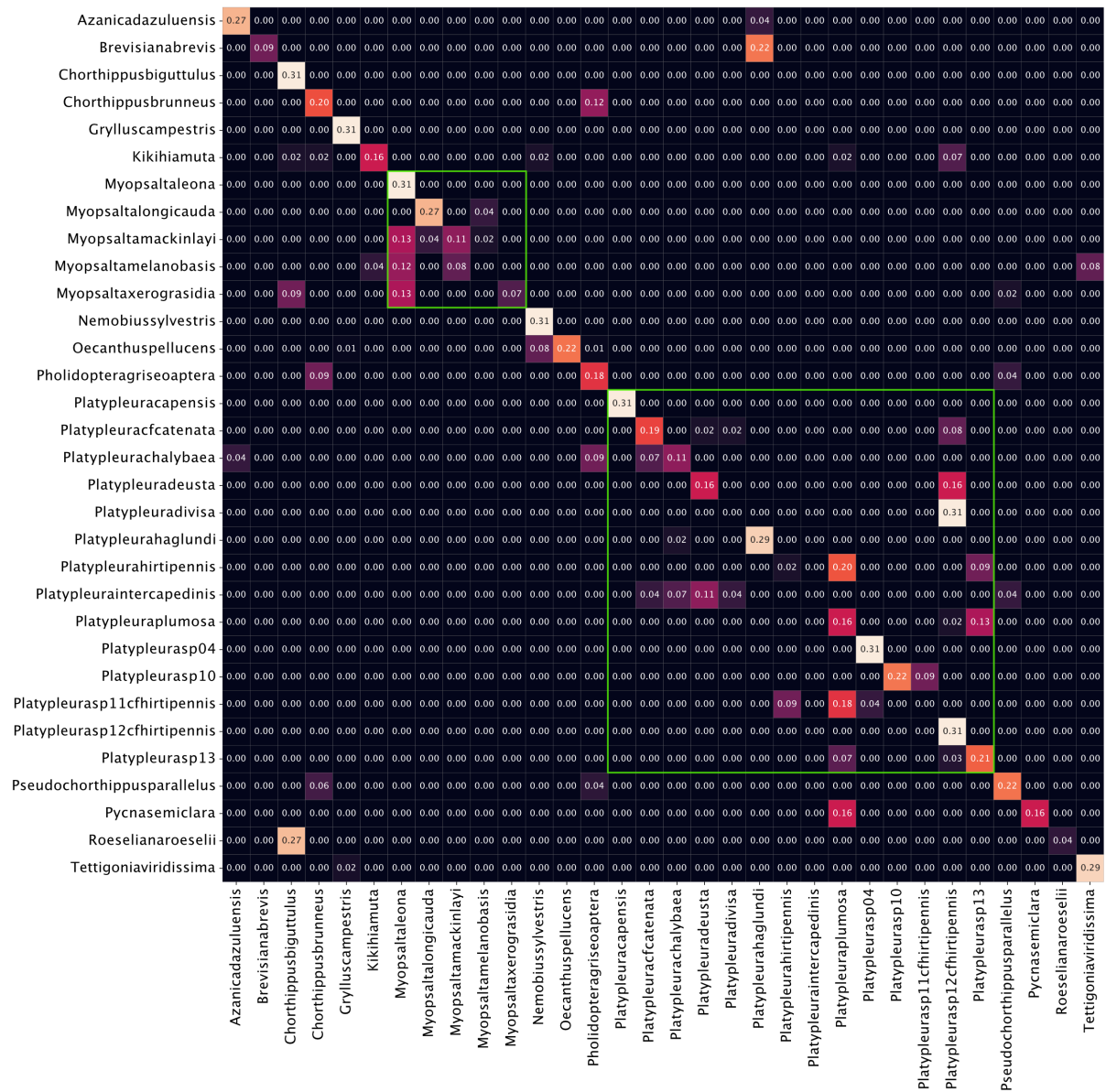

Fig. 7: Confusion matrix showing the classification performance of all classes in the test set using the best iteration of the Mel frontend. The vertical axis displays the true labels of the test files, the horizontal axis shows the predicted labels. Predictions on the diagonal line are correctly identified. Classifications in the two biggest groups *Platypleura* and *Myopsalta* are highlighted since most of the false classifications are between species within these groups.

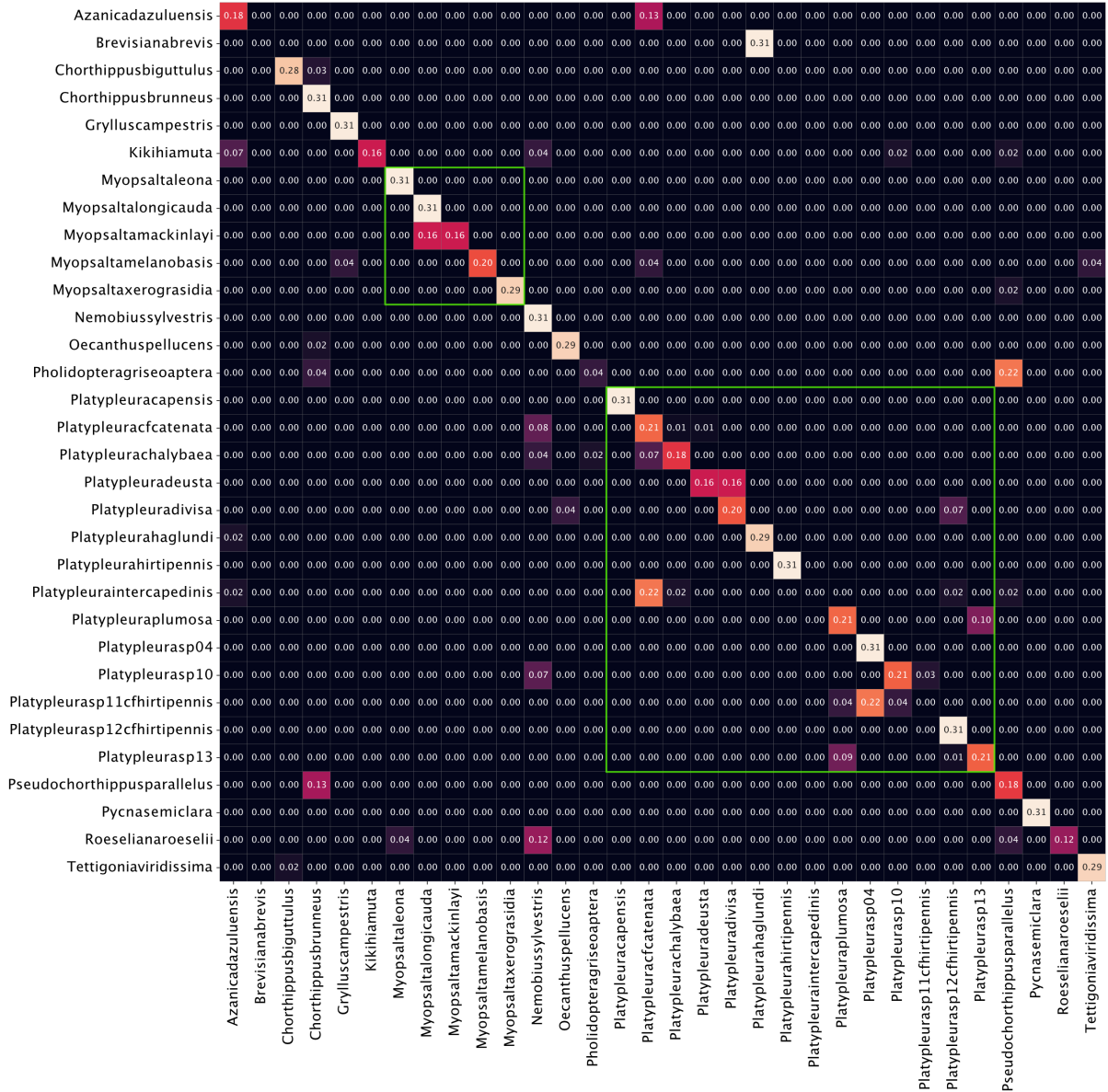

Fig. 8: Confusion matrix showing the classification performance of all classes in the test set using the best iteration of the LEAF frontend. The vertical axis displays the true labels of the test files, the horizontal axis shows the predicted labels. Predictions on the diagonal line are correctly identified. Classifications within the two biggest groups *Platypleura* and *Myopsalta* are highlighted for comparison to the Mel confusion matrix.

The confusion matrix showing the performance of the LEAF frontend on the test set reflects the overall better performance since it displays a clearer diagonal line of accurate classifications, with less incorrect classifications around it (Fig. 8). All test files of the species *Brevisiana brevis* were incorrectly classified as *Platypleura haglundi*. The species *P. intercapedinis* and *P. sp11 cfhirtipennis* were never correctly classified either but confused with different species of the same genus. The concentration of misclassifications in the two largest genera *Myopsalta* and *Platypleura* is much less pronounced compared to the Mel frontend iteration, especially the performance within *Myopsalta* is significantly better (Figs. 7&8).

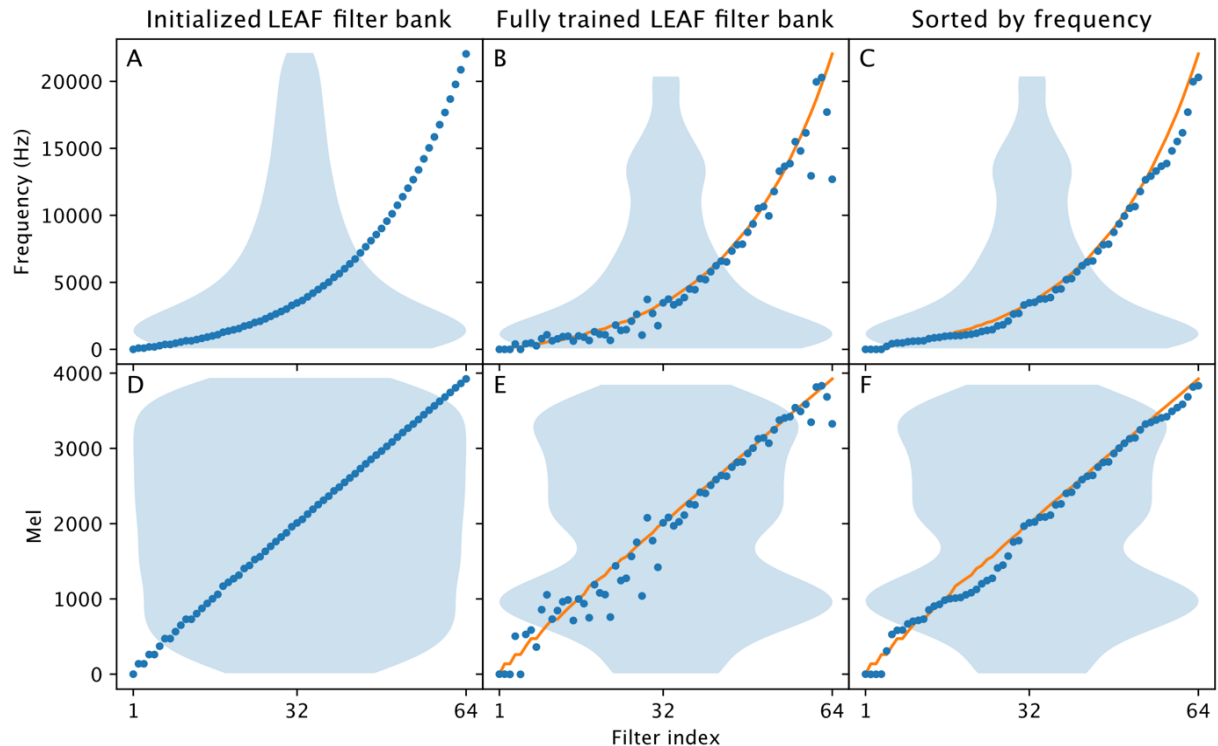

Fig. 9: Center frequencies of all 64 filters used in the best performing LEAF iteration. Plots A and D show the initialization curve before training, which is based on the Mel scale. Plots B and E show the deviation of each filter from their initialized position after training. Plots C and F show the filters sorted by center frequency, and demonstrate the overall coverage of the frequency range, but do not represent the real ordering in the LEAF representations. Violin plots show the density of filters over the frequency spectrum, the orange line shows the initialization curve for comparison.

The filters employed by the LEAF frontend were initialized on a scale closely matched to the Mel scale but were adjusted in center frequency and bandwidth during training (Fig. 9). When sorting the filters by center frequency, the filters still mostly follow the initialization curve (Fig. 9 C&F). Without sorting however, it is clear that many filters were adjusted from their original position (Fig. 9 B&E). Substantial changes in the frequencies of several filters occurred around 2 kHz and above 15 kHz, where some filters were adjusted by up to several kilohertz, especially with the highest filter at initialization being shifted from 22.05 kHz down to approximately 13 kHz (Fig. 9 B). The spatial ordering along the frequency axis is heavily disturbed, since the center frequencies do not steadily increase with increasing filter number, as was the case on the initialized scale (Fig. 9 B&E). This means that in the LEAF outputs matrices, adjacent values on the axis containing frequency information do not necessarily represent adjacent frequency bins, which is usually the case when using hand-crafted representations such as Mel filter banks. Filter density increased around 0.85 kHz (see Fig. 9 D, ≈ 900 Mel) and between roughly 14-15 kHz (Fig. 9 B), but slightly decreased between 18 and 20 kHz (Fig. 9 B) and around 2.4 kHz (see Fig. 9 D, ≈ 1700 Mel). Four filters are located close to zero Mel/kHz after training, leaving a gap up to approximately 500 Mel (≈ 0.4 kHz), where the very lowest insect sound frequencies occur in this dataset (Fig. 9 D).

## Discussion:

The focus of this work was mostly to compare a traditional handcrafted feature extraction method against an adaptive and waveform-based method, while also testing the viability of deep learning methods to classify insect sounds, specifically of Orthoptera and Cicadidae. Using the conventional Mel spectrogram frontend, the best training iteration of the network reached a classification accuracy of 67%. Using the adaptive LEAF frontend, an even better accuracy of 78% was achieved (Table 2). In similar comparisons on more human-centric audio classification tasks (language, emotion, birdsong, music etc.), LEAF outperformed Mel spectrograms on a diverse range of tasks, but not all, and in many cases by much smaller margins than in this comparison [25].

Since the sounds in this comparison are very different in structure and frequency content from human-associated sounds, the difference in performance between LEAF and Mel was expected to be larger than in the previous comparisons, since LEAF can learn a large number of parameters that are completely fixed and not necessarily ideal for the input data when using Mel spectrograms, which is optimized for human sounds. The relevant information in insect sound is largely located in the higher frequency spectrum (above 5 kHz), where Mel spectrograms are more imprecise due to increasingly wider pooling of frequencies. The LEAF frontend adjusted filter center frequencies and bandwidths, as well as compression and time-pooling parameters to better fit the data and reveal details that could be obscured by the Mel frontends fixed parameters (Fig. 9).

The confusion matrices shed some light on where the differences in performance lie between the two approaches (Figs. 7&8). Using the Mel frontend, the majority of incorrect classifications is found between species of the genus *Platypleura*, which represents almost half of the species included in the dataset with 14 out of 32, and in the second largest genus *Myopsalta*, with five species (Table 1). These two groups make up the majority of the species in the dataset and it is therefore more likely for them to contain a majority of the misclassifications. However, the fact that many of their false classifications are within species of the same genus suggests that their sounds could be similar in structure and hard for the network to distinguish. Apparently, the trained parameters of the LEAF frontend led much better performance in these two genera than when using the Mel frontend, since there are less false predictions within these genera and in general.

The overall coverage of filters over the frequency spectrum was not significantly changed during training of the LEAF frontend. When looking at the filter distribution after training, the filters still mostly lie close to the initialization curve that is based on the Mel scale (Fig. 9 C&F). While changes in filter density occurred in some frequency bands, a dramatic shift of for example all filters shifting to higher frequencies or a change to a completely different curve was not observed. When considering the changes of every individual filter however, it is clear that many filters changed position quite significantly, sometimes by several thousand Hertz (Fig. 9 B&E). The ascending order of filter bands along the frequency axis is heavily disturbed after training, meaning that adjacent rows in the LEAF output matrices do not necessarily contain adjacent filter bands. Interestingly, this was not observed in the original paper introducing the LEAF frontend [25] and in a paper improving the performance of the frontend [30]. After training the frontend on the AudioSet dataset [31] and the SpeechCommands dataset [32] at sample rates of 16 kHz, the filters still followed the initialization curve much more closely and the ordering along the frequency axis was conserved in both papers [25,30]. This was interpreted as a demonstration that the Mel scale is a strong initialization curve for these tasks, with the learnable filter parameters in the LEAF frontend mostly providing an opportunity for adapting to a slightly more appropriate frequency range [25,30].

The AudioSet dataset contains many human-centric sounds such as speech and music, as well as a diverse set of environmental sounds, animal sounds and more, with 527 classes and multiple labels per recording [31]. The SpeechCommands dataset contains over 100,000 samples of spoken words [32]. Perhaps this diversity of sounds and classes, as well as the use of a much lower sample rate of 16 kHz [25] constrained the adjustment of filter frequencies compared to the significantly smaller and less diverse dataset used in this comparison. It is also possible that ordering along the frequency axis is more important for classifying sounds that contain defined harmonic structures such as human speech, music, instruments or birdsong. The often noisy and inharmonic sounds produced by Orthoptera and Cicadidae might not require this due to their more uniform and comparably undefined sonic structure over the spectrum.

The order of filters is more disturbed below 5 kHz, which might indicate that ordering in the higher frequency spectrum is more important than in the lower spectrum (Fig. 9 B&E). Or perhaps this is a result of many recordings not containing much audio information in these lower frequencies, since most insect sounds occur in higher ranges. Filters in these lower frequency ranges might have learned to focus on unique frequency bands that are characteristic of a specific species producing relatively lower-frequency sounds. In the higher frequency ranges where sound is present in almost all recordings, a more robust and more frequency-ordered representation might have been advantageous. Filters in the very lowest frequency bands were also adjusted to fit the data. Several filters were adjusted to lie close to zero, creating a gap up to 400 Hz, where the lowest outliers of insect sounds in this dataset occur (Fig. 9 E).

Although the overall distribution of filters over the frequency spectrum was not significantly changed during training, the high occurrence of adjustments and shuffling of individual filters could justify testing different initialization curves than the Mel scale. While this scale seems to be robust and advantageous for classifying human-centric sounds [25], it might not be the ideal initialization curve for insect sounds, especially because the theoretical justifications for the use of the Mel-scale do not apply to the specific characteristics of insect sounds. Perhaps the filter distribution learned in this study is a local optimum that could be reached from the Mel curve as a starting point, but a linear initialization curve might allow the frontend to better adapt to the relevant frequency bands and reach a more global optimum. A more optimal and divergent filter curve using the Mel curve as initialization could also be learned after training for many more epochs than in this study. But testing different initialization curves could allow the frontend to reach a much better and more generalizable filter distribution for insect sounds in a shorter amount of training time, which would be advantageous. One experiment testing a different initialization curve was conducted with randomized center frequency values that were sorted in ascending order [30]. During training, the filter values were adjusted to a more appropriate frequency range for the data, but the overall performance was lower than when using a Mel initialization curve, when tested on the SpeechCommands dataset [30,32]. This, again, shows that the Mel scale is very robust and useful for human sounds, but also that LEAF can learn useful filter distributions even when not initialized on the Mel scale [30]. This further justifies the exploration of alternative initialization scales for usage of the LEAF frontend with non-human sounds.

In the previous studies on the LEAF frontend, the other learnable parameters (pooling, compression and normalization) were adjusted more heavily than the filter parameters and likely contributed more to the improvement in performance than the slight adjustments in filter frequencies [25,30]. A comparison between the LEAF and Mel frontends with disabled Per-Channel Normalization (sPCEN) in the LEAF frontend and disabled learning on the temporal pooling component could be of interest to get a better idea how much the adjustment of the filter parameters alone improves the classification performance. Especially since no Per-Channel Normalization is employed when using the Mel frontend.

The observed difference in performance between the two frontends in this comparison is substantial but could potentially be increased even further by frontend-specific tuning of the model parameters. Some model and regularization hyperparameters were mostly tuned using the Mel spectrogram approach, since the training time is significantly faster. The same parameters were used for both frontends to make the comparison viable, but perhaps slightly different tuning of some parameters or of the training process could reveal an even bigger difference in performance for the LEAF frontend. Ideally, these parameters should be optimized to the same extent for both frontends. It is also possible that a more complex and fine-tuned network could make better use of the potentially more detailed input representations by the LEAF frontend.

To achieve further improvement of classification performance, especially if machine learning methods are going to be implemented in species conservation efforts, larger and more diverse datasets should be the main focus. In this dataset, 32 species were represented, with on average 10 recordings per class. For many species this number was far lower, with some species being represented by only four or five recordings. The fact that this data had to be further split up to create training, validation and test datasets means that a substantial number of species was only trained with examples from two or three recordings. The model can not create a robust representation of these species, since the learning examples likely do not represent the real diversity found in the species.

An opportunity to increase diversity in the dataset as it exists currently could be to manually remove sections with strong noise intrusions in the source datasets, like voice-over commentary. In this work, this was done automatically by extracting only the last ten seconds of many of the files for time reasons. Removing the noise manually would not provide more recordings but would increase the total length of audio material to be split into audio chunks. This way, more examples of natural variation in the sounds would become available for training. But for future use, existing datasets are not sufficient and have to represent all species that occur in the environments where automatic classification methods are going to be deployed and the number of recordings per species has to be increased. Larger and more diverse datasets could especially be of advantage when using LEAF, since it can make use of the additional information to further tweak its parameters and better fit the model, while a non-adaptive frontend can not do this. If datasets with even higher sample rates are going to be used for classification, conventional Mel spectrogram frontends could prove to be even less useful compared to adaptive frontends. Especially for species that produce sounds entirely within the ultrasonic range, which are common in Orthoptera and some Cicadidae [33], the decreasing resolution in high-frequency bands would be increasingly disadvantageous compared to adaptive frontends.

Considering the relatively simple network architecture and small dataset, these results are encouraging for future applications with high potential for further improvements through optimizing model parameters and diversifying datasets. The advantage in performance by using LEAF identifies adaptive frontends as a potentially valuable replacement for approaches with hand-crafted parameters to extract features for insect audio classification. Before these methods can be applied in the real world, datasets need to be increased in size and species diversity, and the networks that are used have to be improved to reach higher overall accuracy. They also need to be integrated with sound-event detection methods to automatically identify relevant clips from longer automatic recordings. This work presents a first step for an important part of the classification network and shows encouraging results and methods for successful future implementations of this technology.

# Appendix

Using MEL

17633 training files at 44100 Hz

for 60 epochs with batch size 14

E1 Training Loss: 3.12, Accuracy: 0.21
E1 Validation Loss: 2.87, Accuracy: 0.30, Patience: 0/8
E2 Training Loss: 2.50, Accuracy: 0.37
E2 Validation Loss: 2.44, Accuracy: 0.43, Patience: 0/8
E3 Training Loss: 2.22, Accuracy: 0.43
E3 Validation Loss: 2.15, Accuracy: 0.41, Patience: 0/8
E4 Training Loss: 2.04, Accuracy: 0.46
E4 Validation Loss: 1.99, Accuracy: 0.47, Patience: 0/8
E5 Training Loss: 1.84, Accuracy: 0.50
E5 Validation Loss: 1.92, Accuracy: 0.48, Patience: 0/8
E6 Training Loss: 1.69, Accuracy: 0.52
E6 Validation Loss: 1.83, Accuracy: 0.55, Patience: 0/8
E7 Training Loss: 1.60, Accuracy: 0.53
E7 Validation Loss: 1.73, Accuracy: 0.57, Patience: 0/8
E8 Training Loss: 1.55, Accuracy: 0.53
E8 Validation Loss: 1.74, Accuracy: 0.54, Patience: 1/8
E9 Training Loss: 1.49, Accuracy: 0.55
E9 Validation Loss: 1.66, Accuracy: 0.56, Patience: 0/8
E10 Training Loss: 1.42, Accuracy: 0.57
E10 Validation Loss: 1.66, Accuracy: 0.56, Patience: 0/8
E11 Training Loss: 1.40, Accuracy: 0.57
E11 Validation Loss: 1.69, Accuracy: 0.54, Patience: 1/8
E12 Training Loss: 1.37, Accuracy: 0.57
E12 Validation Loss: 1.70, Accuracy: 0.56, Patience: 2/8
E13 Training Loss: 1.35, Accuracy: 0.57
E13 Validation Loss: 1.68, Accuracy: 0.57, Patience: 3/8
E14 Training Loss: 1.31, Accuracy: 0.59
E14 Validation Loss: 1.74, Accuracy: 0.53, Patience: 4/8
E15 Training Loss: 1.30, Accuracy: 0.59
E15 Validation Loss: 1.73, Accuracy: 0.57, Patience: 5/8
E16 Training Loss: 1.30, Accuracy: 0.59
E16 Validation Loss: 1.58, Accuracy: 0.61, Patience: 0/8
E17 Training Loss: 1.29, Accuracy: 0.58
E17 Validation Loss: 1.62, Accuracy: 0.60, Patience: 1/8
E18 Training Loss: 1.28, Accuracy: 0.59
E18 Validation Loss: 1.72, Accuracy: 0.63, Patience: 2/8
E19 Training Loss: 1.25, Accuracy: 0.59
E19 Validation Loss: 1.63, Accuracy: 0.64, Patience: 3/8
E20 Training Loss: 1.26, Accuracy: 0.59
E20 Validation Loss: 1.75, Accuracy: 0.61, Patience: 4/8
E21 Training Loss: 1.23, Accuracy: 0.60
E21 Validation Loss: 1.65, Accuracy: 0.65, Patience: 5/8
E22 Training Loss: 1.24, Accuracy: 0.59
E22 Validation Loss: 1.60, Accuracy: 0.60, Patience: 6/8
E23 Training Loss: 1.21, Accuracy: 0.60
E23 Validation Loss: 1.68, Accuracy: 0.58, Patience: 7/8
E24 Training Loss: 1.21, Accuracy: 0.60
E24 Validation Loss: 1.56, Accuracy: 0.64, Patience: 0/8
E25 Training Loss: 1.19, Accuracy: 0.61
E25 Validation Loss: 1.56, Accuracy: 0.61, Patience: 0/8
**E26 Training Loss: 1.23, Accuracy: 0.59**
**E26 Validation Loss: 1.37, Accuracy: 0.65, Patience: 0/8**
E27 Training Loss: 1.20, Accuracy: 0.60
E27 Validation Loss: 1.45, Accuracy: 0.64, Patience: 1/8
E28 Training Loss: 1.19, Accuracy: 0.60
E28 Validation Loss: 1.46, Accuracy: 0.63, Patience: 2/8
E29 Training Loss: 1.18, Accuracy: 0.61
E29 Validation Loss: 1.60, Accuracy: 0.64, Patience: 3/8
E30 Training Loss: 1.19, Accuracy: 0.60
E30 Validation Loss: 1.65, Accuracy: 0.58, Patience: 4/8
E31 Training Loss: 1.18, Accuracy: 0.61
E31 Validation Loss: 1.77, Accuracy: 0.65, Patience: 5/8
E32 Training Loss: 1.18, Accuracy: 0.60
E32 Validation Loss: 1.48, Accuracy: 0.62, Patience: 6/8
E33 Training Loss: 1.16, Accuracy: 0.61
E33 Validation Loss: 1.46, Accuracy: 0.68, Patience: 7/8
E34 Training Loss: 1.19, Accuracy: 0.60
E34 Validation Loss: 1.52, Accuracy: 0.66, Patience: 8/8
Stop Training
Test Accuracy: 0.67, F1 score: 0.56, Recall: 0.58, Precision: 0.64

Suppl. Table 1: Training and Validation scores of the best performing training iteration using the MEL frontend. Training was stopped after 34 epochs after the minimum validation score was reached in epoch 26.

Using LEAF on cuda

17633 training files at 44100 Hz for 60 epochs with batch size 14

E1 Training Loss: 3.08, Accuracy: 0.21
E1 Validation Loss: 2.85, Accuracy: 0.30, Patience: 0/8
E2 Training Loss: 2.47, Accuracy: 0.36
E2 Validation Loss: 2.33, Accuracy: 0.43, Patience: 0/8
E3 Training Loss: 2.18, Accuracy: 0.43
E3 Validation Loss: 2.06, Accuracy: 0.43, Patience: 0/8
E4 Training Loss: 1.97, Accuracy: 0.47
E4 Validation Loss: 1.85, Accuracy: 0.47, Patience: 0/8
E5 Training Loss: 1.79, Accuracy: 0.50
E5 Validation Loss: 1.63, Accuracy: 0.57, Patience: 0/8
E6 Training Loss: 1.65, Accuracy: 0.52
E6 Validation Loss: 1.57, Accuracy: 0.58, Patience: 0/8
E7 Training Loss: 1.58, Accuracy: 0.53
E7 Validation Loss: 1.78, Accuracy: 0.50, Patience: 1/8
E8 Training Loss: 1.53, Accuracy: 0.54
E8 Validation Loss: 1.96, Accuracy: 0.51, Patience: 2/8
E9 Training Loss: 1.51, Accuracy: 0.54
E9 Validation Loss: 1.32, Accuracy: 0.65, Patience: 0/8
E10 Training Loss: 1.48, Accuracy: 0.55
E10 Validation Loss: 1.82, Accuracy: 0.56, Patience: 1/8
E11 Training Loss: 1.48, Accuracy: 0.55
E11 Validation Loss: 1.74, Accuracy: 0.57, Patience: 2/8
E12 Training Loss: 1.43, Accuracy: 0.56
E12 Validation Loss: 1.42, Accuracy: 0.63, Patience: 3/8
E13 Training Loss: 1.41, Accuracy: 0.56
E13 Validation Loss: 1.37, Accuracy: 0.63, Patience: 4/8
E14 Training Loss: 1.38, Accuracy: 0.57
E14 Validation Loss: 2.18, Accuracy: 0.51, Patience: 5/8
E15 Training Loss: 1.39, Accuracy: 0.57
E15 Validation Loss: 1.58, Accuracy: 0.60, Patience: 6/8
E16 Training Loss: 1.40, Accuracy: 0.57
E16 Validation Loss: 1.77, Accuracy: 0.58, Patience: 7/8
E17 Training Loss: 1.38, Accuracy: 0.57
E17 Validation Loss: 1.26, Accuracy: 0.66, Patience: 0/8
E18 Training Loss: 1.37, Accuracy: 0.57
E18 Validation Loss: 1.41, Accuracy: 0.65, Patience: 1/8
E19 Training Loss: 1.37, Accuracy: 0.57
E19 Validation Loss: 1.30, Accuracy: 0.67, Patience: 2/8
E20 Training Loss: 1.37, Accuracy: 0.57
E20 Validation Loss: 1.23, Accuracy: 0.68, Patience: 0/8
E21 Training Loss: 1.34, Accuracy: 0.58
E21 Validation Loss: 1.16, Accuracy: 0.71, Patience: 0/8
E22 Training Loss: 1.35, Accuracy: 0.57
E22 Validation Loss: 1.40, Accuracy: 0.63, Patience: 1/8
E23 Training Loss: 1.31, Accuracy: 0.58
E23 Validation Loss: 1.26, Accuracy: 0.68, Patience: 2/8
E24 Training Loss: 1.29, Accuracy: 0.59
E24 Validation Loss: 1.07, Accuracy: 0.70, Patience: 0/8
E25 Training Loss: 1.30, Accuracy: 0.58
E25 Validation Loss: 1.30, Accuracy: 0.66, Patience: 1/8
E26 Training Loss: 1.28, Accuracy: 0.59
E26 Validation Loss: 1.17, Accuracy: 0.71, Patience: 2/8
E27 Training Loss: 1.27, Accuracy: 0.59
E27 Validation Loss: 1.26, Accuracy: 0.70, Patience: 3/8
E28 Training Loss: 1.28, Accuracy: 0.59
E28 Validation Loss: 1.12, Accuracy: 0.71, Patience: 4/8
E29 Training Loss: 1.30, Accuracy: 0.58
E29 Validation Loss: 1.08, Accuracy: 0.71, Patience: 5/8
**E30 Training Loss: 1.27, Accuracy: 0.59**
**E30 Validation Loss: 1.00, Accuracy: 0.76, Patience: 0/8**
E31 Training Loss: 1.26, Accuracy: 0.59
E31 Validation Loss: 1.03, Accuracy: 0.73, Patience: 1/8
E32 Training Loss: 1.25, Accuracy: 0.59
E32 Validation Loss: 1.03, Accuracy: 0.74, Patience: 2/8
E33 Training Loss: 1.25, Accuracy: 0.59
E33 Validation Loss: 1.08, Accuracy: 0.72, Patience: 3/8
E34 Training Loss: 1.26, Accuracy: 0.59
E34 Validation Loss: 1.17, Accuracy: 0.72, Patience: 4/8
E35 Training Loss: 1.24, Accuracy: 0.60
E35 Validation Loss: 1.21, Accuracy: 0.67, Patience: 5/8
E36 Training Loss: 1.24, Accuracy: 0.60
E36 Validation Loss: 1.07, Accuracy: 0.72, Patience: 6/8
E37 Training Loss: 1.23, Accuracy: 0.60
E37 Validation Loss: 1.16, Accuracy: 0.70, Patience: 7/8
E38 Training Loss: 1.25, Accuracy: 0.59
E38 Validation Loss: 1.13, Accuracy: 0.71, Patience: 8/8
Stop Training
Test Accuracy: 0.78, F1 score: 0.69, Recall: 0.70, Precision: 0.73

Suppl. Table 2: Training and Validation scores of the best performing training iteration using the LEAF frontend. Training was stopped after 38 epochs after the minimum validation score was reached in epoch 30.

```
AudioClassifier:
 leaf: Leaf()
   _complex_conv: GaborConv1d()
     constraint: GaborConstraint()
   _activation: SquaredModulus()
     _pool: AvgPool1d(kernel_size=(2,), stride=(2,), padding=(0,))
   _pooling: GaussianLowPass()
   _compression: PCENLayer()
     ema: ExponentialMovingAverage()
 conv1: Conv2d(1, 8, kernel_size=(5, 5), stride=(2, 2), padding=(2, 2))
 relu1: ReLU()
 bn1: BatchNorm2d(8, eps=1e-05, momentum=0.1, affine=True, track_running_stats=True)
 conv2: Conv2d(8, 16, kernel_size=(3, 3), stride=(2, 2), padding=(1, 1))
 relu2: ReLU()
 bn2: BatchNorm2d(16, eps=1e-05, momentum=0.1, affine=True, track_running_stats=True)
 conv3: Conv2d(16, 32, kernel_size=(3, 3), stride=(2, 2), padding=(1, 1))
 relu3: ReLU()
 bn3: BatchNorm2d(32, eps=1e-05, momentum=0.1, affine=True, track_running_stats=True)
 conv4: Conv2d(32, 64, kernel_size=(3, 3), stride=(2, 2), padding=(1, 1))
 relu4: ReLU()
 bn4: BatchNorm2d(64, eps=1e-05, momentum=0.1, affine=True, track_running_stats=True)
 ap: AdaptiveAvgPool2d(output_size=1)
 lin: Linear(in_features=64, out_features=32, bias=True)
 dropout: Dropout(p=0.4, inplace=False)
```

Suppl. Table 3: Full model specification used in the experiment with 27344 trainable parameters. The data was not passed though the LEAF frontend when the Mel frontend was used and vice versa.